\documentclass[aps,pra,reprint,groupedaddress]{revtex4-1}
\usepackage[utf8]{inputenc}

\usepackage{amsmath}
\usepackage{fixmath}

\usepackage{soul}

\usepackage{graphicx}

\begin{document}

\title{Detailed study of a transverse field Zeeman slower}

\author{D. Ben Ali}
\author{T. Badr}
\author{T. Brézillon}
\author{R. Dubessy}
\author{H. Perrin}
\author{A. Perrin}
\email[]{aurelien.perrin@univ-paris13.fr}
\affiliation{Laboratoire de physique des lasers, CNRS UMR 7538, Université Paris 13, Sorbonne Paris Cité, F-93430, Villetaneuse, France.}

\date{\today}

\begin{abstract}
We present a thorough analysis of a Zeeman slower for sodium atoms made of permanent magnets in a Halbach configuration. Due to the orientation of the magnetic field, the polarisation of the slowing laser beam cannot be purely circular leading to optical leakages into dark states. To circumvent this effect, we propose an atomic state preparation stage able to significantly increase the performances of the Zeeman slower. After a careful theoretical analysis of the problem, we experimentally implement an optical pumping stage leading to an increase of the magneto-optical trap loading rate by 3.5. Such method is easy to set up and could be extended to other Zeeman slower architectures.
\end{abstract}


\maketitle

\section{Introduction}

Zeeman slowers are a popular technology widely used to decelerate thermal beams of atoms to velocities where they can be laser cooled and trapped. The various designs that have been proposed and implemented since its first realisation~\cite{Phillips1982} all rely on the radiation pressure from a contra-propagating laser beam. In order to compensate for the Doppler frequency shift and maintain the laser beam at resonance with an electronic transition of the atom, a static and spatially varying magnetic field is used to induce an opposite Zeeman frequency shift. The shape of this magnetic field is chosen to fit the velocity change of the atoms along their trajectories.

The orientation of the magnetic field inside the apparatus defines a preferred quantization axis which in turn determines the polarisation of laser beams. In the different Zeeman slower implementations, the magnetic field is either colinear~\cite{Phillips1982,Streed2006,Ovchinnikov2012,Lebedev2014,Krzyzewski2014,Zhang2015,Wang2015,Zhang2016,Hill2014} or orthogonal to the atom propagation axis~\cite{Hill2014,Bagayev2001,Ovchinnikov2007,Cheiney2011,Lamporesi2013a}. In the first case, pure circular polarisation of the laser beam is achievable allowing for an atom-light interaction limited to a cycling transition, within an effective two-level system. In the other case, a pure circular polarisation of the laser beam is not possible and only the fraction of the laser field which is properly polarised contributes to a cycling transition. The interaction of the atom with other polarisation components of the laser beam may lead to optical leakages into dark states. Appropriate strategies are required to avoid this effect and preserve the performances of the Zeeman slower.

Here we present a thorough analysis of a transverse magnetic field Zeeman slower based on permanent magnets in a Halbach configuration~\cite{Halbach1980}. Such design has the advantage to be compact and easy to install and remove without the need for electric power nor water cooling. It also allows for very smooth magnetic field profiles with excellent transverse homogeneity and low stray magnetic fields. It has originally been introduced for $^{87}$Rb atoms~\cite{Cheiney2011} and we show how to adapt it to $^{23}$Na atoms taking into account their specific level structure. Thanks to numerical simulations based on the resolution of optical Bloch equations, we are able to precisely identify the origin of the leakage into dark states, due to the orientation of the magnetic field in this configuration. This allows us to propose an adequate response based on a proper atomic state preparation upstream from the Zeeman slower. We finally present experimental measurements demonstrating the performances of our implementation and conclude with general considerations about Zeeman slower technologies.

The organisation of the paper is as follows. In Section~\ref{section2} we present our experimental design of a Zeeman slower in a Halbach configuration dedicated to sodium atoms. This is the opportunity to introduce a Matlab routine\footnote{The Matlab code with a few example scripts is given as supplementary materials.\label{Note1}} able to optimise the design of any Zeeman slower in a Halbach configuration. In Section~\ref{section3} we present results of numerical simulations of the atom-light interaction upstream from the Zeeman slower. Depending on the orientation of the magnetic field and on the laser beam polarisation, we show how the atoms tend to accumulate in the dark hyperfine state of the sodium ground state. We then propose a method to circumvent this issue relying on an optical pumping stage with two laser beams. As we show, an atom optically pumped into the right magnetic substate doesn't suffer from significant leakages into dark states once inside the Zeeman slower. In Section~\ref{section4} we present the experimental performances of our Zeeman slower design and stress how the optical pumping stage significantly enhances the flux of decelerated atoms feeding a magneto-optical trap (MOT). We finally discuss our results and give general conclusions in section~\ref{section5}.

\section{Experimental design}\label{section2}

Like all alkali atoms, the sodium ground state presents an hyperfine structure~\cite{Steck2010} and cycling transitions are only available on the $D_2$ line. Our Zeeman slower design follows the same architecture than the one described in~\cite{Cheiney2011} adapting its shape to fit sodium atoms specificities. As in~\cite{Cheiney2011}, we work with an increasing magnetic field configuration. As a result, the laser beam addresses the $|F=2,m_F=-2\rangle$ to  $|F'=3,m_{F'}=-3\rangle$ transition, where $F$ and $F'$ represent respectively the total angular momentum of the ground and excited states and $m_F$ and $m_{F'}$ their projections onto the quantization axis. In the rest of the paper we label $|S\rangle=|F=2,m_F=-2\rangle$ the ground state of this slowing transition.

\subsection{Magnetic field profile within a simple model}\label{simplemodel}

Modelling the cycling transition as a two level system and using a semi-classical approach, the force induced by the radiation pressure onto the atoms writes~\cite{Metcalf1999}:
\begin{eqnarray}\label{force}
\mathbf{F}=-\eta\frac{\Gamma}{2} \hbar k \mathbf{u}_z
\end{eqnarray}
where $\hbar$ is the Planck constant, $k$ the light wave vector, $\Gamma$ the natural line width of the transition, $\mathbf{u}_z$ the unit vector along the Zeeman slower axis $z$ and 
\begin{eqnarray}
\eta = \frac{s_0}{1+s_0+4\delta^2/\Gamma^2}.
\end{eqnarray}
Here $\delta=\omega_{\rm laser}-\omega_{\rm atom}$ is the laser beam detuning to the atomic transition, $s_0 = I/I_{\rm sat}$ the resonant saturation parameter, $I$ the laser beam intensity and $I_{\rm sat}$ the saturation intensity of the transition.

Assuming that the atom propagates along the $z$ axis against the laser beam, in the presence of a magnetic field $B$ and taking into account the Doppler frequency shift due to the atom velocity $\mathbf{v}=v\mathbf{u}_z$, the detuning writes
\begin{eqnarray}
\delta &= \delta_0+kv-\frac{\mu_B B}{\hbar}(g_{F'}m_{F'}-g_Fm_F) \\
 &\simeq \delta_0+kv+\frac{\mu_B B}{\hbar}\label{detuning}
\end{eqnarray}
where $\delta_0$ is the detuning for an atom at rest and in the absence of magnetic field, $\mu_B$ the Bohr magneton and $g_F$ and $g_{F'}$ the Land\'e factors of the hyperfine states.

If $\delta$ remains constant along the atom propagation, the radiation pressure force is also constant and energy conservation implies the following relation for the velocity $v(z)$ at position $z$
\begin{eqnarray}
v(z)^2=v(0)^2-\eta\frac{\hbar k}{m}\Gamma z. \label{energy}
\end{eqnarray}

Combining Eq.~\eqref{detuning} and \eqref{energy} sets the model profile of the magnetic field along the atom propagation~\cite{Phillips1982}:
\begin{eqnarray}
B(z) = B(0) + \Delta B \left(1-\sqrt{1-\frac{z}{L}}\right)\label{magneticfield}
\end{eqnarray}
with $B(0)=\hbar(\delta-\delta_0-kv(0))/\mu_B$, $\Delta B = \hbar k v(0)/\mu_B$ and $L=mv(0)^2/(\eta\hbar k \Gamma)$.

\subsection{Choice of the experimental parameters}\label{choice}

According to Eq.~\eqref{magneticfield}, in order to set the magnetic field shape for experimental implementation, the values of $\Delta B$, $B(0)$ and $L$ -- or equivalently $\eta$ -- have to be defined. 

The choice of $\Delta B$ determines the initial velocity $v(0)$ which corresponds to the largest velocity of the atoms that will be decelerated in the Zeeman slower. Ideally, in order to maximize the flux of decelerated atoms, the latter has to be as large as possible compared to the peak value of the velocity distribution in the thermal effusive beam of atoms originating from the oven, which writes~\cite{Ramsey1986}
\begin{eqnarray}\label{distrib}
P(v)=\left(\frac{m}{k_B T}\right)^2\frac{v^3}{2}\exp\left(-\frac{m v^2}{2 k_B T}\right).
\end{eqnarray}
In the experiment, the oven has a temperature of $T=570$~K. Then, imposing $v(0)=950$~m/s allows for the slowing of about $64\%$ of the atomic flux. This leads to $\Delta B=1152$~G which is on the order of the highest amplitude of the magnetic field achievable with rare-earth magnets in a Halbach configuration taking into account typical experimental constraints.

For a given value of the laser beam intensity $I$, the parameter $\eta$ only depends on $|\delta|$ and is bounded from above by $s_0/(1+s_0)<1$. This sets a lower limit on the value of $L$. To compensate for possible local imperfections of the magnetic field achieved experimentally, it is safer to design its profile with a strictly positive value for $|\delta|$. Setting $\eta=0.52$ leads to $L=0.96$~m. This in turn imposes a minimum value for the laser intensity $I_{\rm tot}>I_{\rm min}\simeq13.6$~mW/cm$^2$. Here, the laser polarisation is chosen linear and orthogonal to the direction of the magnetic field, which maximises the projection onto the relevant $\sigma^-$ polarisation addressing the slowing transition, i.e $I=I_{\rm tot}/2$.

\begin{figure}[htb]
\centering\includegraphics[width=8.5cm]{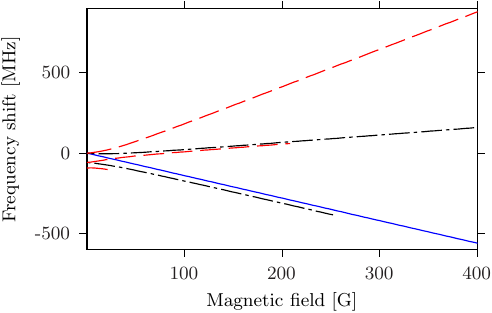}
\caption{\label{figureCrossings}
Frequency shifts of the transitions from state $|S\rangle$ as a function of the magnetic field. The solid blue, dashed-dotted black and dashed red curves correspond respectively to $\sigma^-$, $\pi$ and $\sigma^+$ transitions. The only allowed $\sigma^-$ transition has a constant electric dipole matrix element $d_{\rm max}$ across the entire magnetic field range. For $\pi$ and $\sigma^+$ transitions, we follow each transition up to the magnetic field where the associated electric dipole matrix element falls below $0.05 d_{\rm max}$.
}
\end{figure}

Once $\Delta B$ and $L$ are set, the value of $B(0)$ is arbitrary but must remain positive to avoid a zero magnetic field region which cannot be easily designed using permanent magnets. This in turn determines the detuning of the laser beam $\delta_0$ since $|\delta|\ll\delta_0$. A value of $B(0)=300$~G leads to $\delta_0/(2\pi)\simeq-2$~GHz which allows to almost suppress the effect of the laser beam onto the magneto-optical trap where the atoms are captured after deceleration. Moreover, with such a value for $B(0)$, electronic level crossings within the excited state of the atoms are also avoided inside the Zeeman slower~\cite{Steck2010}, as illustrated in figure~\ref{figureCrossings}. Above 300~G, there is a single isolated $\sigma^-$ transition, which justifies the use of an effective two-level model to describe the deceleration process~\cite{Joffe1993}. The residual leakage into dark states due to the other, far-detuned, transitions is studied quantitatively in section~\ref{section3}.

In order to define the output velocity of the Zeeman slower, one could in principle slightly shorten the length of the magnetic field profile. Since for $z\simeq L$ the slope of the magnetic field profile is almost vertical, see Eq.~\eqref{magneticfield}, it is actually more efficient experimentally to slightly adjust the detuning $\delta_0$ to optimize it.

\subsection{Experimental implementation}

The magnetic field of a Zeeman slower should ideally be transversally homogeneous across the atomic beam and should not leak outside the apparatus. Only few technological designs are actually able to produce magnetic fields respecting these two constraints. The most common one relies on current-carrying wires wound around the atomic beam. It readily produces a transversely homogeneous magnetic field pointing along the axis of the solenoid. However it often represents a large amount of metal that needs to be carefully wound in order to produce a smooth magnetic field profile. Moreover, once built it cannot be easily disassembled and limits baking possibilities for the vacuum chamber. Finally, in order to reach high magnetic field values, it requires a large current circulating in the coils~\cite{Bell2010}, representing hundreds of watts of power that needs to be dissipated.

\begin{figure*}[htb]
\centering\includegraphics[width=15cm]{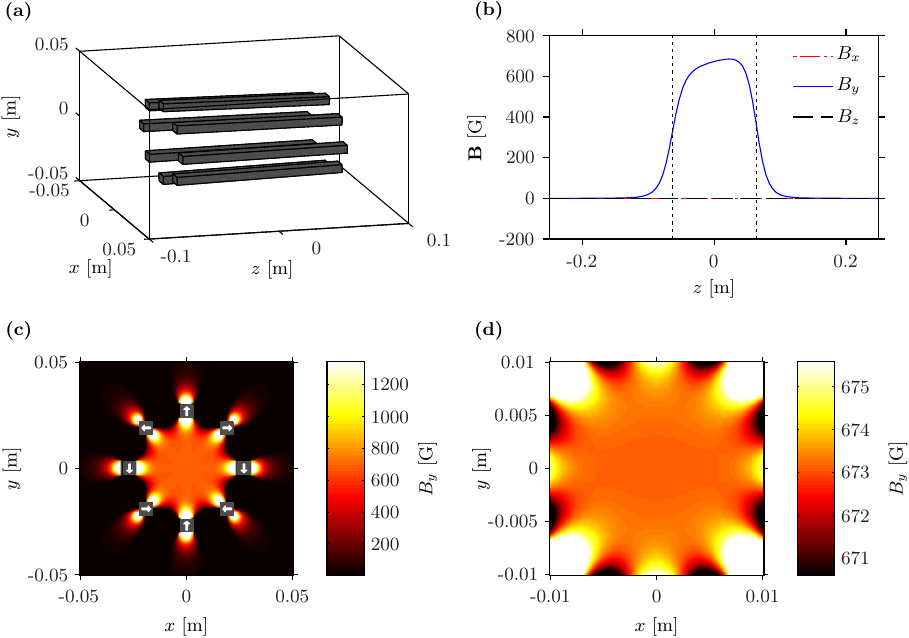}
\caption{\label{figure2}
(a) Three-dimensional representation of a Halbach ring made of 8 parallelepipedic magnets of size $6\times 6\times 128$~mm and remanent field $B_R=10.8$~kG. The mean diameter of the ring is 5.4~cm and its angle with respect to the symmetry axis of the assembly is 0.97$^{\circ}$ (b) Amplitude of the three components of the magnetic field $\mathbf{B}$ along the symmetry axis of the Halbach ring. The results are based on the exact formula for the magnetic field induced by a cuboid magnet presented in the supplementary information of Ref.~\cite{Cheiney2011}. The vertical dotted lines indicate the length of one magnet of the assembly. (c) Amplitude of $B_y$ in the $x-y$ plane for $z=0$. The orientation of the magnetization of each magnet of the Halbach ring is indicated with a white arrow.(d) Zoom on the central part of (c).}
\end{figure*}

Alternatively it is possible to use permanent magnets in order to mimic a Halbach configuration~\cite{Halbach1980}. Here, the orientation of the magnetization of the magnets is such that the magnetic field is homogenous within the Halbach ring and quickly falls to zero outside~\cite{Cheiney2011}. Fig.~\ref{figure2}~(a) shows an example involving eight NdFeB parallelepipedic magnets\footnote{HKCM, part number: Q128x06x06Zn-30SH} of size $6\times 6\times 128$~mm$^3$ and remanent field $B_R=10.8$~kG. The resulting magnetic field, deduced from numerical calculations, points along the $y$ axis and is relatively homogeneous transversely as shown in Fig.~\ref{figure2}~(b),(c) and (d). In order to tune the amplitude of the magnetic field along the $z$-axis, one can slightly tilt the magnets individually with respect to the symmetry axis of the ring. Such assembly can be seen as a building block for Zeeman slowers.

\begin{figure*}[htb]
\centering\includegraphics[width=15cm]{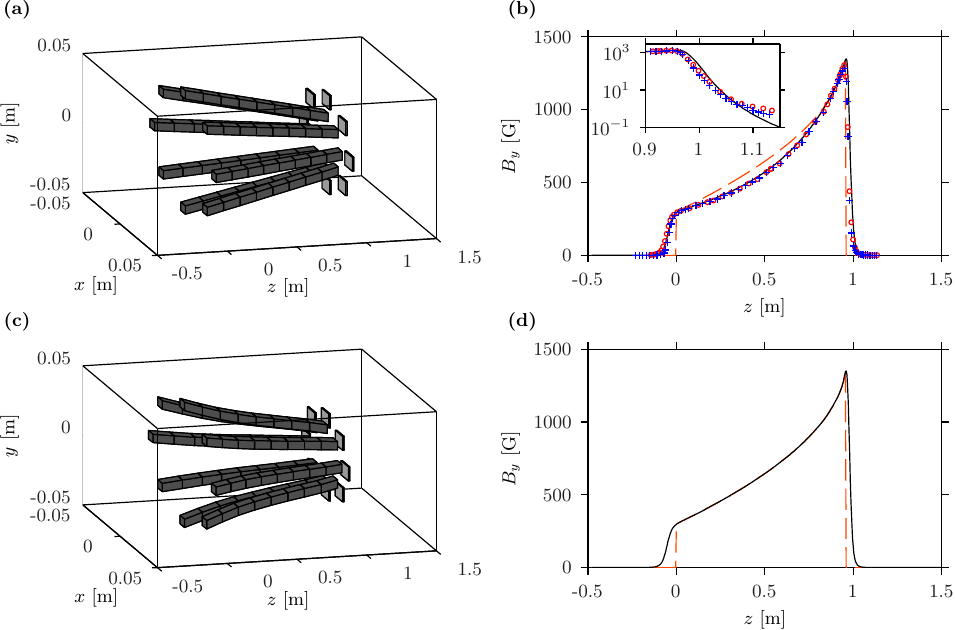}
\caption{\label{figure3}
(a) Three-dimensional representation of the experimental implementation of our Zeeman slower. It is fabricated with nine consecutive Halbach rings made of NdFeB permanent magnets. The first eight rings are built with parallelepipedic 30SH grade magnets of size $6\times 6\times 128$~mm$^3$. The diameter of the first ring is 80~mm and the angle of the magnets with respect to the symmetry axis of the assembly is $1.25^\circ$. The last ring is made of 10~mm-side cubic magnets of N35 grade. Its diameter is 61~mm. (b) Longitudinal profile of the $y$-component of the magnetic field deduced from Eq.~\eqref{magneticfield} (red dashed line), obtained from exact calculations taking into account the geometry of the magnets shown in (a) (black line) or from experimental measurements with (blue crosses) or without (red circles) a magnetic shield surrounding the magnets assembly. The inset shows in log scale a zoom of the last centimetres of the magnetic field profile. (c) Three-dimensional representation of a Zeeman slower where the magnets in the different rings can have different angles with respect to the symmetry axis of the assembly. (d) Comparison of the model profile of the magnetic field (red dashed line) and the one obtained from exact calculations of the field produced by magnets assembly shown in (c)  (black solid line).}
\end{figure*}

Combining eight successive Halbach rings as shown in Fig.~\ref{figure3}~(a) and (c) allows to reproduce the model profile of Eq.~\eqref{magneticfield} taking into account the parameters introduced in section~\ref{choice} (see Fig.~\ref{figure3}~(b) and (d)). Since the shape of the model profile is quite steep toward the end, an additional ring made of 10~mm-side cubic magnets of N35 grade ($B_R=11.7$~kG) is added at the end of the structure~\cite{Cheiney2011}. In order to adjust the geometry of the Zeeman slower shown in Fig.~\ref{figure3}~(a), we have optimised by hand the values of the diameters of the different Halbach rings, their position along the $z$-axis and the angle of the magnets with respect to the symmetry axis of the ring. Since a design where the angles of the magnets are all different would require to be able to hold them individually,  we have opted for a simpler mechanical solution close to Ref.~\cite{Cheiney2011} implementation where all the magnets apart from the last ring share the same angle in a guide. This leads to Fig.~\ref{figure3}~(a) and (b) where our experimental measurements show very good agreement with the theoretical expectation.

It is worth noting that since the magnetic field produced by each magnet decreases within its length on each end (see also Fig.~\ref{figure2}~(b)), the total length of the magnets assembly is larger than the extension of the magnetic field profile. Moreover, as the model profile becomes extremely steep toward the output of the Zeeman slower, reproducing experimentally its shape with accuracy, and in particular its maximum, is beyond reach.
In our experimental realisation, the measured value of the magnetic field maximum is 100~G lower than the theoretical one. This implies that the experimental values for the capture velocity and the output velocity will be smaller than the theoretical ones.
It is not an issue as they can be readily fine-tuned by changing the laser detuning. The experimental optimum for the MOT loading rate is actually $\delta_0/(2\pi)=-1885$~MHz.

Adding a magnetic shield made of a 1~mm-thick soft iron sheet around the permanent magnets assembly improves the fall off of the magnetic field outside the Zeeman slowing region (see inset Fig.~\ref{figure3}~(b)). At the position of the MOT, the residual magnetic field drops from 1~G to 0.5~G thanks to the shielding. The complete assembly is also fairly easy to install and remove on the experimental setup and the resulting magnetic field is much smoother than what is achievable with wound wires.

Setting as constraints that the different Halbach rings must lie next to each other along the $z$-axis and that the diameter of the rings must decrease continuously at the exception of the last ring, and letting the angles of each magnet be free parameters, we have been able to check using a Matlab optimisation routine~\cite{Note1} that it is possible to almost perfectly reproduce the model profile of the magnetic field as shown in Fig.~\ref{figure3}~(d). The routine is versatile and once the number of Halbach rings is fixed, allows for the optimisation of any parameter of the magnets geometry: their sizes, the diameter of the different rings and their angle with respect to the symmetry axis of the Zeeman slower. It is also possible to optimize the parameters of the model profile, for example $\eta$, which can be interesting if the length of the magnets is fixed since it controls the length of the Zeeman slower. We have made a few tests with different model profiles and the optimisation routine converges typically after a few tenths of seconds. The field resulting from the optimisation is in good agreement with the model profile. The relative error is typically smaller than $0.1\%$ when the angle of each magnet is optimised independently (see Fig.~\ref{figure3}~(c)).

\section{Numerical simulations}\label{section3}

\subsection{Atom trajectories}\label{traj}

\begin{figure*}[htb]
\centering\includegraphics[width=15cm]{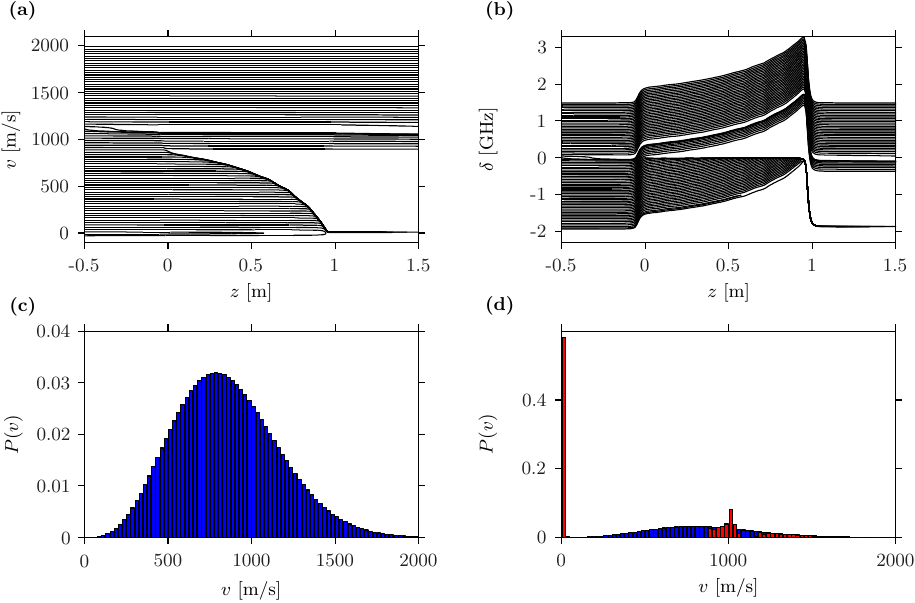}
\caption{\label{figure4}
(a) Simulation of the trajectories in velocity space for atoms with different initial velocities inside the actual Zeeman slower implemented in the experiment (see Fig.~\ref{figure3}~(b) for the magnetic field profile). (b) Behaviour of the detuning $\delta$ for the atomic trajectories shown in (a). (c) Initial velocity distribution at the output of the oven, following Eq.~\eqref{distrib}. (d) Velocity distribution at the output of the oven (blue bars) and at the output of the Zeeman slower (red bars). The bin size for histograms (c) and (d) is 25~m/s.}
\end{figure*}

Since the experimental implementation of the Zeeman slower doesn't produce a magnetic field profile which perfectly reproduces the one deduced from Eq.~\eqref{magneticfield}, it is interesting to simulate the classical trajectories of atoms belonging to different velocity classes to check how the efficiency of the assembly is affected. From Eq.~\eqref{force} we deduce the acceleration of each atom, which depends on its velocity and on the amplitude of the magnetic field at the position of the atom. We can then solve the equation of motion and infer the trajectory of the atoms in velocity space within the two-level model. Taking into account the experimental magnetic field profile shown in Fig.~\ref{figure3}~(b) and setting the laser intensity to $I_{\rm tot}=87$~mW/cm$^2$ and the detuning to $\delta_0/(2\pi)=-1885$~MHz as in the experiment, see section~\ref{section4}, we obtain the results presented in Fig.~\ref{figure4}~(a). Up to a velocity slightly smaller than 900~m/s, the atoms decelerate inside the Zeeman slower to reach a velocity close to 13~m/s. Atoms starting with an initial velocity smaller than this latter value are ultimately decelerated to a zero velocity before being pushed back. Atoms with an initial velocity close to 1150~m/s start at resonance with the laser beam and are therefore slightly slowed down before entering the Zeeman slower. All the other velocity classes are unaffected. From Fig.~\ref{figure4}~(b), we observe that the detuning $\delta$ of all the decelerated atoms increases up to a value on the order of $-\Gamma$ where it locks, the force remaining constant for the rest of the trajectory.  Taking into account the initial velocity distribution of the atoms at the output of the oven following Eq.~\eqref{distrib}, depicted in Fig.~\ref{figure4}~(c), we obtain in Fig.~\ref{figure4}~(d) the final velocity distribution of the atoms at the output of the Zeeman slower. About $60\%$ of the initial atoms are hence slowed down inside the magnets assembly.

\subsection{Optical pumping}\label{optic_pump}

As mentioned above it is not possible to define a purely circular polarisation for the slowing laser beam in a Zeeman slower using permanent magnets in a Halbach configuration, because the resulting magnetic field is perpendicular to the beam propagation axis. The best choice, a linear polarisation orthogonal to the magnetic field, can be written as the sum with equal weights of $\sigma^-$ and $\sigma^+$ polarisations.
Hence an atom interacting with the slowing beam can be optically pumped into  dark states, which cannot be slowed down.
In fact, for the increasing field Zeeman slower we built, only the ground state $|S\rangle$ can be efficiently slowed.
In order to evaluate the fraction of slowed atoms, we perform a numerical calculation to compute precisely the fraction of atoms that remain in this state during their propagation from the output of the oven to the magneto-optical trap chamber.
We start with the low magnetic field section, i.e. the first 50 centimetres of the ballistic flight, from the output of the oven to the input of the Zeeman slower, where significant optical pumping occurs. We then show how this depumping effect can be mitigated by introducing a transverse repumping section. Finally we focus on the dynamics inside the Zeeman slower where optical pumping is very limited thanks to the large Zeeman splitting, see Fig.~\ref{figureCrossings}.

We proceed as follows.
We use the framework of the optical Bloch equations (OBE) to describe the interaction of the internal atomic degrees of freedom with the laser beams~\cite{Steck2010,Cohen-Tannoudji1998}.
Because we are mainly interested in optical pumping phenomena that occur at long time scales and result in population transfer, we reduce the full OBE system to rate equations describing only the time evolution of the atomic populations.
We perform this reduction by considering that the atomic coherences follow adiabatically the populations.
This approximation is not appropriate to describe the short time dynamics but we checked that it leads to an accurate determination of the long term dynamics and correctly recovers the steady state.
Moreover it is quite simple to implement numerically and results in a dramatic speed up of the computation besides reducing the memory requirements ($576\times576$ matrices to describe the full sodium D$_2$ line within the OBE, $24\times24$ matrices for the rate equation model).

Starting with an atom with a longitudinal velocity $v$ at the output of the oven, we compute the evolution of its internal state populations due to the interaction with the detuned Zeeman slower beam, starting from a thermal equilibrium equipartition between the ground state sublevels.
Between the output of the oven and the input of the Zeeman slower itself the atom propagates for 50~cm in the presence of a low magnetic field, mostly due to the Earth magnetic field.
Therefore the atom absorbs on average very few photons and we can neglect the radiation pressure force, which in turn allows to simplify the numerical treatment of the dynamics.
However the depumping effects are not negligible, as shown below.
We typically follow the evolution of the internal atomic states by monitoring the populations every centimetre along the atom trajectory.

\begin{figure}[htb]
\centering\includegraphics[width=8.5cm]{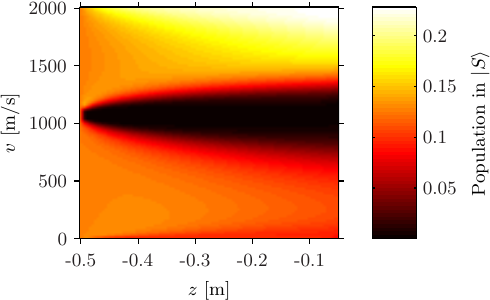}
\caption{\label{figureEBO1}
Population in state $|S\rangle$ for different initial velocities as a function of the distance travelled by the atoms from the output of the oven at $z=-0.5$~m to the entrance of the Zeeman slower at $z=0$~m.
Velocity dependent depumping along the propagation is due to the presence of the counter-propagating linearly polarised Zeeman slower laser beam, see text for detail.
}
\end{figure}

Figure~\ref{figureEBO1} shows the computed evolution of the population in level $|S\rangle$ along the propagation for different initial velocities close to the output of the oven, with the same values of the detuning $\delta_0$ and the total laser intensity $I_{\rm tot}$ as those used in section~\ref{traj}.
For velocity classes close to the resonance, a strong optical pumping towards states of the $F=1$ hyperfine manifold occurs.
As the state $|S\rangle$ belongs to the $F=2$ hyperfine manifold at low magnetic fields, only a small fraction of the atoms can be slowed by the Zeeman slower.
For the theoretical velocity distribution of Fig.~\ref{figure4}~(c), we compute that only $3.7~\%$ of the atoms enter the Zeeman slower in the $|S\rangle$ state.

\begin{figure}[htb]
\centering\includegraphics[width=8.5cm]{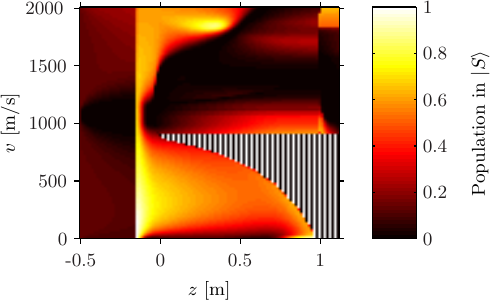}
\caption{\label{figureEBO2}
Population in state $|S\rangle$ for different initial velocities as a function of the distance travelled by the atoms starting from the output of the oven (at $z=-0.5$~m).
At $z=-0.15$~m a repumping section is included, see text for detail.
Beyond this point, the atoms enter the increasing magnetic field region of the Zeeman slower and the depumping dynamics is strongly affected. The vertically hatched pattern indicates the phase space domain where atoms in the $|S\rangle$ state are efficiently slowed.
}
\end{figure}

In order to mitigate this optical pumping effect we propose to pump the atoms back into the $|S\rangle$ state using a transverse beam, with two frequencies tuned to the $|F=1\rangle\to|F^\prime=2\rangle$ and to the $|F=2\rangle\to|F^\prime=2\rangle$ transitions with purely circular polarisation.
Because the lasers are perpendicular to the atomic beam and can be resonant for all velocity classes, this process can be made very efficient, as can be seen in figure~\ref{figureEBO2} at position $z=-0.15$~m.
Unfortunately, due to geometrical constraints this optical pumping stage cannot be done immediately at the entrance of the Zeeman slower and the atoms propagate again in a low magnetic field for about $15$~cm. We thus expect that $31~\%$ of the atoms enter the Zeeman slower in the relevant state.

Finally we consider the evolution of the atomic populations inside the Zeeman slower, in the presence of the increasing magnetic field, under the assumption that the atomic populations follow adiabatically the magnetic field changes.
More specifically we model the continuous magnetic field inside the Zeeman slower by discrete steps of constant magnetic field, with an increment of $3$~G, and compute the atomic population evolution for this particular field value, over the appropriate distance (determined mainly by the local magnetic field gradient).
The output populations are then fed to the next segment.
As the magnetic field increases, the matrix elements of the electric dipole between the different atomic states are modified as the system progressively enters the Paschen-Back regime and the two circular components of the linearly polarised Zeeman slower laser beam become more and more asymmetrically detuned with respect to the allowed transitions, see figure~\ref{figureCrossings}.
At sufficiently high magnetic field these two effects ensure that the slowing transition is almost closed such that the atoms are slowed, provided they reach the `slowing area' in the $|S\rangle$ internal state.
Here we numerically investigate to which extent this affirmation is true and in particular we quantify the effect of the intermediate field area where level crossings could favour optical pumping into non slowed states and reduce the efficiency of the Zeeman slower.

Figure~\ref{figureEBO2} shows the computed evolution of the population in level $|S\rangle$ along the propagation for different initial velocities, from the oven to the magneto-optical trap chamber, including the repumping section using a transverse laser beam at position $z=-0.15$~cm.

After an initial depumping section, as in figure~\ref{figureEBO1}, the atoms are repumped into the level $|S\rangle$ by the transverse beams, resulting in 31~\% of the atoms entering the Zeeman slower in state $|S\rangle$. The populations subsequently evolve in the presence of the detuned Zeeman laser beam and the increasing magnetic field, resulting in a more complex behaviour.
First, above the Zeeman slower capture velocity $v_c\simeq 900$~m/s, the population in level $|S\rangle$ is depleted inside the Zeeman slower.
As the Zeeman slower is not designed to slow these atoms this is not an issue.
Then, for velocities smaller than $v_c$, the depumping mechanism becomes less efficient as the field increases and about $28~\%$ of the atoms end in the slowable state at the position where the Zeeman laser becomes resonant (indicated by the vertically hatched pattern).
Therefore the additional depumping due to this section is negligible in our setup.
For the lowest velocities the depumping is significant inside the Zeeman slower, even at high magnetic fields, due to the long interaction times.

Our model predicts an increase of the slow atom flux at the output of the Zeeman slower, with about four times more atoms when the repumping section is present.
This figure is computed by comparing the number of atoms in the $|S\rangle$ state for each velocity class at the position inside the Zeeman slower where it is resonant with the slowing laser, convoluted by the velocity distribution, respectively with and without the transverse repumping section described above.

\section{Experimental performances}\label{section4}

In order to set up an atomic state preparation on the experimental apparatus, we have installed a laser beam propagating horizontally along the horizontal $x$-axis orthogonal to the atomic beam of sodium, roughly 15~cm before the entrance of the Zeeman slower. It corresponds to the closest position where a window is available on the set-up to let the beam pass through the vacuum chamber. The laser is circularly polarised and its frequency is set close to the $|F=2,m_F=-1\rangle\rightarrow|F'=2,m_{F'}=-2\rangle$ transition. The total power in the beam is 25~mW for a $1/e^2$-diameter of 8~mm. We use an electro-optical modulator to produce sidebands at 1.771~GHz from the carrier frequency. They both represent about $20\%$ of the total laser power. In order to split the Zeeman sublevels of the atoms in this region where only the Earth magnetic field is present, we apply a homogeneous magnetic field pointing in the direction of the propagation axis of the laser and produced by two 20-turn coils in a Helmholtz configuration. Its amplitude is on the order of 1~G. While the carrier optically pumps the $|F=2\rangle$ atoms into the  $|F=2,m_F=-2\rangle$ state, the blue sideband is used to repump the atoms from the $|F=1\rangle$ to the $|F=2\rangle$ states.

The slowing beam itself possesses a waist of 0.5~cm for an intensity of $I_{\rm tot}=87$~mW/cm$^2$. Its detuning is $\delta_0/(2\pi)=-1885$~MHz and it is linearly polarised along the $x$ axis which means that only half of the total power of the beam contributes to the slowing transition.

\begin{figure}[htb]
\centering\includegraphics[width=8.5cm]{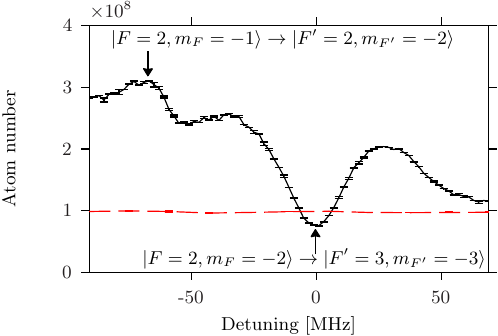}
\caption{\label{figure6}
Dependence with the optical pumping laser detuning of the number of atoms captured in the magneto-optical trap after 3~s of loading with the atomic state preparation (black line, see text) or without (red line). The detuning is zero when the laser is resonant with the $|F=2,m_F=-2\rangle\rightarrow|F'=3,m_{F'}=-3\rangle$ transition. The error bars come from the standard deviation computed over two realisations of the experiment.
}
\end{figure}

Loading a magneto-optical trap (MOT) from the slowed atomic beam for 3~s allows to test the performances of the atomic state preparation. The dependence of the captured atom number with the laser detuning is shown in figure~\ref{figure6}\footnote{The absolute performance of the MOT loading rate is mainly determined by the parameters of the MOT beams (sizes, intensities,...), the temperature of the sodium oven and the geometry of the vacuum chamber (aperture of the oven output, distance between the oven output and MOT chamber, apertures of the differential pumping stages,...) and the amplitude of stray magnetic fields in the MOT chamber. We have been able to increase the absolute MOT loading rate up to $10^9$ atoms/s by optimizing some of these parameters. We have observed that the relative increase in the MOT loading rate due to the optical pumping stage is independent of the absolute MOT loading rate.}. The pump laser intensity has been optimized here such that increasing it further does not change the final populations anymore. When the pumping laser is resonant with the $|F=2,m_F=-2\rangle\rightarrow|F'=3,m_{F'}=-3\rangle$ transition, the atoms end up cycling over the transition, exchanging a large number of photons with the laser which increases significantly the divergence and the deflection of the atomic beam, degrading in turn the performances of the MOT loading. This is no more the case when the laser is resonant with the $|F=2,m_F=-1\rangle\rightarrow|F'=2,m_{F'}=-2\rangle$ transition since the $|F=2,m_F=-2\rangle$ state is in this case a dark state and all the atoms get optically pumped into it. It is worth noting that since the magnetic field direction is different in the optical pumping area (along the $x$ axis) compare to the slowing region inside the Zeeman slower (along the $y$ axis), we assume the spin of the atoms to adiabatically follow the local direction of the magnetic field along their propagation. 

In order to compare the efficiency of the atomic state preparation with the simulations of section~\ref{optic_pump}, we have measured the dependence of the MOT atom number with its loading time. Comparing the slope of the atom number with and without the atomic state preparation, we measure a factor of 3.5 increase in the atomic loading. This result agrees well with the factor of four deduced from the simulations. We attribute the small discrepancy to the probable over-estimation of the efficiency of the repumping section in the computation as we did not include the spatially varying intensity of the laser in the numerics and modelled the real gaussian shaped laser beams by top-hat beam profiles.

\begin{figure*}[htb]
\centering\includegraphics[width=15cm]{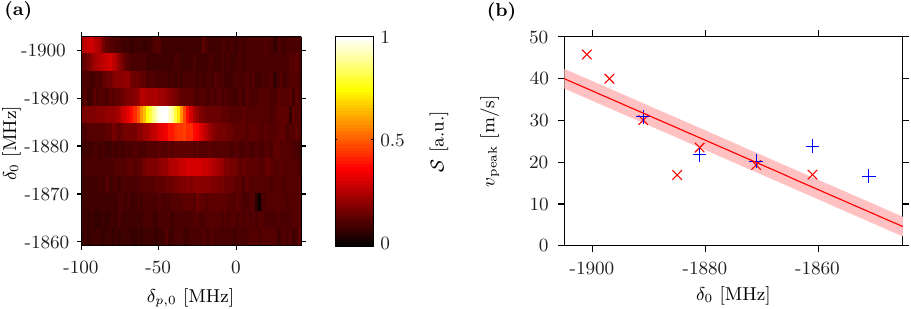}
\caption{\label{figure10}
(a) Fluorescence signal $\mathcal{S}$ in arbitrary units as a fonction of $\delta_{p,0}$ and $\delta_0$ induced by a probe beam crossing the atomic beam at $-45^\circ$. (b) Dependence of $v_{\rm peak}$ as a fonction of $\delta_0$ for a probe beam crossing the atomic beam at respectively $-45^\circ$ (red crosses), $45^\circ$ (blue plus signs). The red line is a linear fit of the results. The shaded area around the line represents the uncertainty of the fit.}
\end{figure*} 

We have also measured the peak velocity of the slowed beam by monitoring the fluorescence signal $\mathcal{S}$ induced by a probe beam crossing the atomic beam at $\pm45^\circ$~\cite{Phillips1982}. This signal depends on the velocity of the atoms through Doppler effect. For a single atom with longitudinal velocity $\mathbf{v}$, the probe beam detuning $\delta_p$ writes
\begin{eqnarray}
\delta_p = \delta_{p,0}-\mathbf{k}_p\cdot\mathbf{v}=\delta_{p,0}\pm \frac{k_p v}{\sqrt{2}}
\end{eqnarray}
with $\delta_{p,0}$ its detuning for an atom at rest and $k_p$ the modulus of its wavevector. We present in Fig.~\ref{figure10}~(a) the dependence of $\mathcal{S}$ with $\delta_{p,0}$ and the detuning $\delta_0$ of the slowing laser beam. For a given $\delta_0$, the observed peak in the fluorescence signal $\mathcal{S}$ corresponds to the peak velocity $v_{\rm peak}$ of the velocity distribution of the slowed atoms : $\delta_{p,0}=\mp k_pv_{\rm peak}/\sqrt{2}$. The observed velocity width is limited by the natural width of the transition. As shown in Fig.~\ref{figure10}~(b), the peak velocity of the slowed atomic beam depends linearly with $\delta_0$ as expected from Eq.~\eqref{detuning} if we assume that the velocity of the atoms remains constant after exiting the Zeeman slower. This is a fair assumption from the results of the simulations shown in Fig.~\ref{figure4}~(a). For a detuning of -1885~MHz we find $v_{\rm peak}=28(3)$~m/s, slightly higher than the estimation of section~\ref{traj}. This discrepancy can be explained by a 1~\% uncertainty on the magnetic field measurement near the end of the Zeeman slower, where the magnetic field is high and varies rapidly, see figure~\ref{figure3}.

\section{Discussion}\label{section5}

We now turn to more general considerations on the available methods to mitigate the optical pumping effects in the transverse field Zeeman slower considered in this paper.
This discussion is enabled by the accurate modelling of the optical pumping dynamics that we have developed.
We will focus on the increasing field Zeeman slower for sodium described in the present work, for which we have shown that a transverse repumping strategy is efficient, resulting in up to $28~\%$ of the total atomic flux being slowed.

Let us first consider the case of a perfectly polarised slowing light, which could be achieved by a more complex arrangement of permanent magnets resulting in a longitudinal magnetic field (see for example Ref.~\cite{Ovchinnikov2012}). We compute that $32~\%$ of the flux would be slowed, due to optical depumping occurring during the initial population redistribution in the $|F=2\rangle$ manifold, before the entrance of the Zeeman slower. Again, a transverse repumping section is needed to reach the theoretical limit of $60~\%$ of the flux set by the estimated velocity capture.

The authors of Ref.~\cite{Cheiney2011} have shown that another solution was to use a longitudinal repumping laser, co-propagating with the slowing light, with a frequency either fixed or rapidly modulated over the whole Doppler width.
We only computed the former case as the latter would be much more involved. Briefly we extended our OBE description by adding the contribution of the second laser and instead of performing a rotating wave approximation we performed a Floquet expansion truncated at second order. This allows to consider single photon absorption as well as two photon transitions but neglects higher order phenomena. We find that for a repumping laser tuned to the $|F=1\rangle\to|F^\prime=2\rangle$ transition (at low magnetic field) about $19~\%$ of the total flux is slowed.
From our computations it seems that it is particularly efficient inside the Zeeman slower, where it pumps back to the $|S\rangle$ state a set of velocity classes locally resonant at a given magnetic field.
We may expect that the result could be even better with a modulated repumping frequency.
However this strategy is very demanding in terms of laser sources and probably necessitates an independent laser system for the repumping beam, which can be very costly, especially for sodium atoms.

To conclude, we have presented an efficient optical pumping method able to significantly increase the performance of a transverse field Zeeman slower made of permanent magnets in a Halbach configuration. We have experimentally demonstrated an increase by a factor of 3.5 of the atomic loading rate in the MOT. This figure could be even improved if the distance between the optical pumping stage and the entrance of the Zeeman slower could be further reduced. We believe that this technique could be extended to many different Zeeman slower architectures in order to optimise their efficiencies and in turn reduce the MOT loading time or limit the total atomic flux to improve the background vacuum pressure.

\begin{acknowledgments}
This work has been supported by the Region Ile-de-France in the framework of DIM IFRAF (Institut Francilien de Recherche sur les Atomes Froids) and by the ANR Project No. 11-PDOC-021-01. LPL is UMR 7538 of CNRS and Paris 13 University.
\end{acknowledgments}

\bibliography{biblio}

\end{document}